\documentclass[10pt, conference, letterpaper]{IEEEtran}

\AtBeginDocument{%
  \providecommand\BibTeX{{%
    \normalfont B\kern-0.5em{\scshape i\kern-0.25em b}\kern-0.8em\TeX}}}

\usepackage{graphicx}
\usepackage{textcomp}
\usepackage{hyperref}
\usepackage{xcolor}
\usepackage{balance}
\usepackage{subcaption}
\usepackage{multirow}
\usepackage{booktabs}
\hypersetup{
    colorlinks=true, 
    linkcolor=black, 
    filecolor=black,
    urlcolor=black,
    citecolor=black, 
}

\usepackage{url}
\usepackage{soul}
\usepackage{makecell}
\usepackage{float}
\usepackage{amsmath}
\usepackage{algorithm} 
\usepackage{packages/algpseudocodex}
\usepackage{amsthm} 
\usepackage{amssymb}
\usepackage{authblk}
\usepackage{lipsum}
\usepackage{enumitem}

\makeatletter
\renewcommand{\ALG@beginalgorithmic}{\small}
\renewcommand{\algorithmicindent}{1em} 
\makeatother

\newcommand\blfootnote[1]{%
  \begingroup
  \renewcommand\thefootnote{}\footnote{#1}%
  \addtocounter{footnote}{-1}%
  \endgroup
}

\DeclareMathOperator*{\argmax}{arg\,max}

\title{BAROC: Concealing Packet Losses in LSNs with Bimodal Behavior Awareness for Livecast Ingestion}

\author{
Haoyuan Zhao\textsuperscript{\dag}, 
Jianxin Shi\textsuperscript{\ddag}\textsuperscript{\dag}, 
Guanzhen Wu\textsuperscript{\dag}, 
Hao Fang\textsuperscript{\dag},
Yi Ching Chou\textsuperscript{\dag},\\
Long Chen\textsuperscript{\dag},
Feng Wang\textsuperscript{\textparagraph},
Jiangchuan Liu\textsuperscript{\dag}\\
\textsuperscript{\dag}School of Computing Science, Simon Fraser University, Canada\\
\textsuperscript{\ddag}College of Cryptology and Cyber Science, Nankai University, China\\
\textsuperscript{\textparagraph}Department of Computer and Information Science, University of Mississippi, USA\\
Emails: hza127@sfu.ca, shijianxin@mail.nankai.edu.cn, \{gwa52, fanghaof, ycchou\}@sfu.ca, \\
longchen.cs@ieee.org, fwang@cs.olemiss.edu, jcliu@sfu.ca
\vspace{-2em}
}

\begin{document}
\maketitle
\begin{abstract}

The advent of Low-Earth Orbit satellite networks (LSNs), exemplified by initiatives like \emph{Starlink}, \emph{OneWeb} and \emph{Kuiper}, has ushered in a new era of ``Internet from Space" global connectivity. Recent studies have shown that LSNs are capable of providing unprecedented download capacity and low latency to support Livecast viewing. However, Livecast ingestion still faces significant challenges, such as limited uplink capacity, bandwidth degradation, and the burst of packet loss due to frequent satellite reallocations, which cause previous recovery and adaptive solutions to be inferior under this new scenario. In this paper, we conduct an in-depth measurement study dedicated to understanding the implications of satellite reallocations, which reveals that the network status during reallocations with network anomalies exhibits a different distribution, leading to bimodal behaviors on the overall network performance. Motivated by this finding, we propose BAROC, a framework that can effectively conceal burst packet losses by combining a novel proposed MTP-Informer with bimodal behavior awareness during satellite reallocation. BAROC enhances video QoE on the server side by addressing the above challenges and jointly determining the optimal video encoding and recovery parameters. Our extensive evaluation shows that BAROC outperforms other video delivery recovery approaches, achieving an average PSNR improvement of $1.95$ dB and a maximum of $3.44$ dB, along with enhancements in frame rate and parity packet utilization. Additionally, a comprehensive ablation study is conducted to assess the effectiveness of MTP-Informer and components in BAROC.

\end{abstract}

\blfootnote{This research was supported by an NSERC Discovery Grant and a MITACS Accelerate Cluster Grant. The corresponding author is Jiangchuan Liu.}
\blfootnote{This is the preprint version of the paper accepted to IEEE INFOCOM 2025.}

\section{Introduction}
\label{sec:intro}

Since \emph{Starlink} first started the commercial service of Low-Earth Orbit (LEO) satellite constellation, both industry and academia have witnessed its impressive network performance and noticed its great potential of achieving global Internet coverage in the upcoming era of 6G and beyond. After establishing saturated coverage in most urban areas, the leader operator, \emph{Starlink}, \emph{OneWeb}, and \emph{Kuiper}, expanded its service to include land mobility, maritime, and aviation to realize the proposed Space–Air–Ground network. One of the primary applications benefiting from the expansion of service areas is Livecast, where streamers are typically limited to urban areas because of the need for stable, high-capacity networks. With the expansion of LEO Satellite Networks (LSNs), anytime and anywhere real-time video task applications, such as outdoor Livecasts, wilderness rescues, and surveillance, are becoming increasingly feasible.

\begin{figure}[t]
     \centering
     \includegraphics[width=\linewidth]{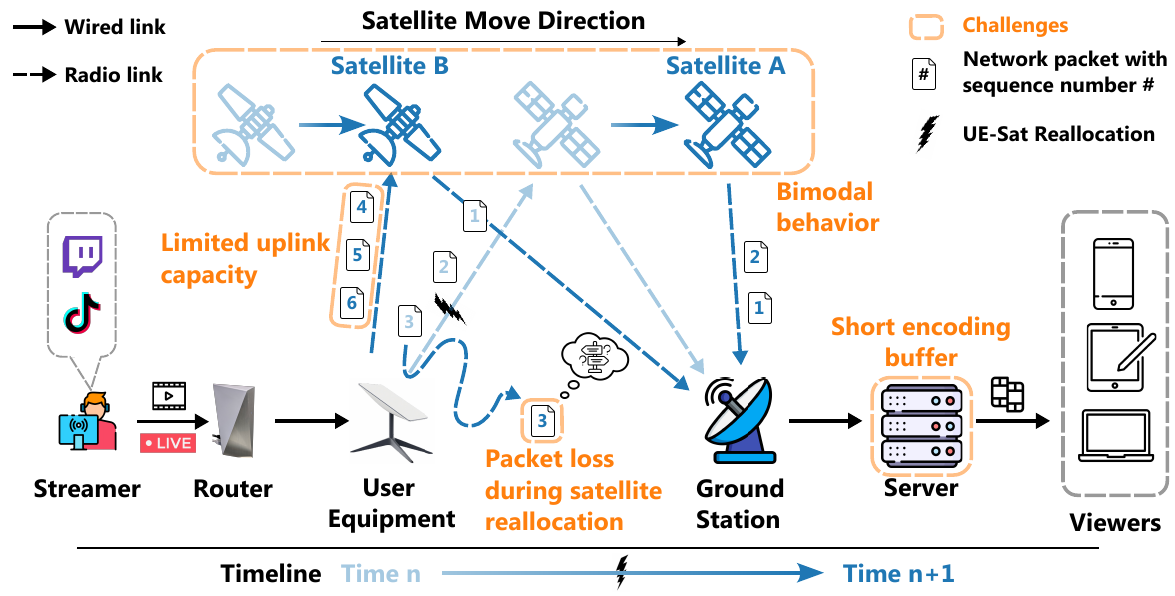}
     \caption{Overview of Livecast ingestion over LSNs and the relevant challenges.}
     \label{fig:packet_loss_by_reallocation}
     \vspace{-7mm}
\end{figure}

In a typical Livecast scenario, streamers first capture high-quality video content, upload it to the cloud, and then distribute it to viewers through Content Delivery Networks. Despite the high downlink capacity of LSNs, which can reach around $200$Mbps and latency of approximately $50$ms, capable of real-time, high-quality video viewing \cite{mohan2024multifaceted, ma2023network, michel2022first, kassem2022browser, zhao2023realtime}, Livecast ingestion still encounter significant challenges due to substantial performance disparities between LSN's uplink and downlink, as well as higher delivery requirements. Firstly, since each LEO satellite can only cover a small area and moves quickly, it is needed to switch and relay the signal to the next satellite to maintain a ``seamless" connection with the user equipment (UE) (as shown in Fig. \ref{fig:packet_loss_by_reallocation}). \textit{Frequent satellite reallocations lead to bursts of packet loss and sudden bandwidth degradation} \cite{mohan2024multifaceted, tanveer2023making, cao2023satcp}, increasing network unpredictability and degrading Livecast ingestion performance. While initial efforts have been made to mitigate such impacts for the downlink of Livecast \cite{cao2023satcp, wu2024accelerating, li2024satguard, zhao2024low, fang2024robust}, the more challenging issues of real-time ingestion caused by satellite reallocation remains largely unexplored. 

Secondly, due to the limited antenna size and power supply employed in the UE, the UE has less transmitting signal power compared to the satellites. For instance, \textit{the uplink capacity is $15$x smaller than the downlink capacity, which is only around $15$Mbps}. Thirdly, from the application perspective, the encoding buffer on the server side is much smaller than the video content buffer on the viewer's side, meaning any packet loss in the uplink will quickly halt the decoding and distribution pipeline on the server, affecting all downstream viewers. Therefore, strategies like slowing down playback speed to extend buffer duration \cite{zhao2024low, fang2024robust} are not feasible for the streamer as the content creator. 

In this work, we conduct an in-depth measurement study dedicated to understanding the implications of satellite reallocations, which reveals that \textit{the network status during reallocations with network anomalies exhibits a different distribution, leading to bimodal behaviors} on the overall network performance. Motivated by this finding, we propose a bimodal behavior aware loss concealing (BAROC) framework for Livecast ingestion. To address packet loss during Livecast, the most common methods are packet- or frame-level recovery \cite{emara2022ivory, lee2022r, hu2021lightfec, ray2022prism}, as re-transmission is generally impractical due to Livecast latency constraints. These methods usually rely on an accurate predictor to determine the appropriate ratio of redundant (a.k.a parity) packets or require sufficient computation power and time for high-quality frame recovery using Neural encoders. Due to the aforementioned challenges inherently from LSNs, these prerequisites are difficult to meet. To meet the constrained time requirement from the application perspective, an efficient recovery method, such as the forward error correction (FEC) mechanism is recommended. Yet, considering the inclusion of extra parity packets for recovering, and the limited uplink capacity mentioned in LSNs, an accurate LSN-specific network predictor and a more efficient video quality scheduler are also required to maintain both high video quality and packet recovery ratio.

To this end, in the BAROC framework, we propose an enhanced Transformer predictor \cite{vaswani2017attention, zhou2021informer}, Multi-Task Probabilistic Informer (MTP-Informer), and with a video quality scheduler that jointly resolves the distribution convolution problem introduced by bimodal behavior and bitrate variance. This framework maximizes the video quality on the server side by jointly adjusting the FEC ratio (the ratio of parity packets) and video encoding parameters on the streamer side considering both satellite reallocation, limited uplink capacity, and variance of video encoded bitrate. 

To adapt to the bimodal behavior of the LSNs, the \textit{MTP-Informer models the probability distribution of the bimodal network metrics, rather than relying solely on single-point observations}. It predicts the future probability distribution, providing a more informative prediction for later video scheduling. This approach makes the BAROC aware of less frequent packet loss bursts and also more resilient to inaccurate forecasts. We then formulate a convolution calculation problem that combines the probability distribution of bimodal behaviors and the distribution arising from bitrate variance and resolves with the proposed video quality scheduler.

To evaluate the performance of BAROC, we compare the proposed solution with other state-of-the-art video delivery recovery methods using real-world LSN traces with multiple videos incorporating typical Livecast scenarios. Our emulated experiment shows that BAROC demonstrates improvements in terms of Peak signal-to-noise ratio (PSNR), frame rate, parity packet utilization ratio, and recovery ratio. In detail, BAROC achieves an average PSNR improvement of $1.95$ dB, with a maximum of $3.44$ dB, and an average increase of $13.54\%$ in parity packet utilization. Additionally, we conduct an ablation study to solely evaluate the effectiveness of components within the BAROC framework. The contributions of this paper can be summarized as follows.

\begin{itemize}[left=0pt]
    \item We conduct a measurement study focused on the impact of satellite reallocations, uncovering the underlying bimodal behaviors. The fluctuating packet loss and bandwidth during these reallocations highlight the need for a novel Livecast ingestion framework tailored to LSNs. (\S \ref{sec:background})
    \item We propose the enhanced MTP-Informer specifically for LSNs, which predicts the probability distribution of network metrics based on historical satellite reallocation effect and bimodal behavior. (\S \ref{sec:method})
    \item We formulate the distribution convolution problem and introduce a video quality scheduler that jointly determines the optimal video quality and FEC parameters based on the predicted bandwidth and packet loss distributions. (\S \ref{sec:formulation})
    \item We evaluate the performance of BAROC with other state-of-the-art real-time video delivery recovery methods (R-FEC \cite{lee2022r}, LightFEC \cite{hu2021lightfec}, and FBRA \cite{nagy2014congestion}) and demonstrate its superiority in terms of multiple evaluation metrics. (\S \ref{sec:evalution})
\end{itemize}
\section{Background and Motivation}
\label{sec:background}

\textbf{LEO networks and satellite reallocations}. For most LSN operators, such as \emph{Starlink} and \emph{OneWeb}, their commercial UE is outfitted with a single-phased array antenna for sending and receiving signals. During UE-Satellite (UE-Sat) link rescheduling, the UE requests a new spot beam from the incoming satellite and releases the current one upon confirmation. During this beam switch, packets in the UE-Sat link, such as packet 3 in Fig. \ref{fig:packet_loss_by_reallocation}, may be lost due to delays in routing table updates or resource reconfiguration \cite{mohan2024multifaceted, cao2023satcp, maattanen20195g}.

In practice, we observe that bursts of lost packets can affect tens or even hundreds of packets sequentially, which will also significantly degrade the bandwidth. Both documentation and measurement work \cite{mohan2024multifaceted, tanveer2023making, WinNT} have confirmed that the \emph{Starlink} employs a global controller for UE-Sat link reallocation schedule, where the UE-Sat link will be rescheduled every $15$ second, and specifically at $12$th, $27$th, $42$th, and $57$th second in each minute for all UEs. \textit{While the reallocation scheduling is known, the actual impact of reallocation on end-to-end network performance remains ambiguous}. The reallocation effect can exhibit either significant network degradation or minimal visible impact most of the time, resulting in bimodal behaviors, as demonstrated by our measurement study to be discussed next. Furthermore, compared to reallocation in cellular or WiFi networks, satellite reallocation has more pronounced consequences. For example, in LSNs, the satellite is mobile while the user (UE) remains relatively static, causing all users to experience periodic reallocation effects during usage. For cellular or WiFi networks, reallocations only affect users who are moving away from the current access point.

\begin{figure}[t]
     \centering
     \includegraphics[width=\linewidth]{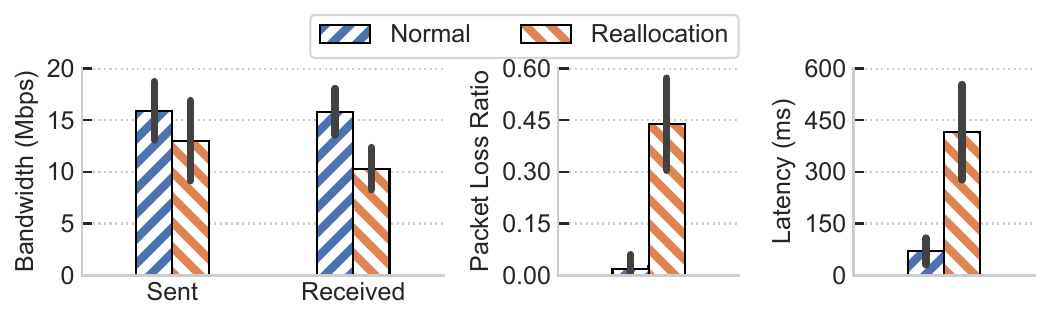}
     \caption{Performance measurement of Livecast over LSNs.}
     \label{fig:reallocation_compare}
     \vspace{-5mm}
\end{figure}

\begin{figure}[t]
     \centering
     \includegraphics[width=\linewidth]{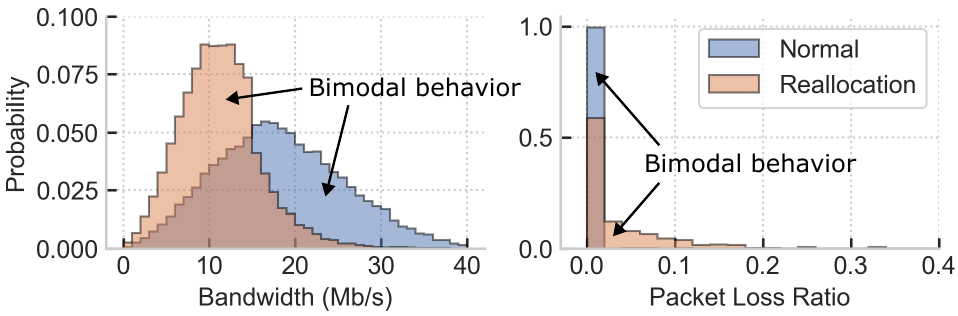}
     \caption{The probability distribution of network metrics during reallocation and normal period.}
     \label{fig:dist_compare}
     \vspace{-5mm}
\end{figure}

\textbf{Measurements on bimodal behaviors in LSNs.} Prior works \cite{mohan2024multifaceted, ma2023network, michel2022first, kassem2022browser} have presented a comprehensive network evaluation on LSNs, our focus is specifically on examining LSNs' performance degradation during reallocation, particularly concerning bandwidth, latency, and packet loss ratio. For a typical Livecast scenario, the recommended maximum packet loss ratio is less than $2\%$ according to \emph{Zoom's} User guideline\footnote{\url{https://support.zoom.com/hc/zh/article?id=zm_kb&sysparm_article=KB0070504}}. Thus, we define a period of 1-second as network anomaly if we observe the average packet loss ratio is larger than $2\%$. 

During our week-long measurement (detailed measurement setup in \S \ref{sec:experiment_setup}), $30.73\%$ of reallocation periods were marked as network anomaly, compared to only $4.32\%$ normal period marked as network anomaly. Fig. \ref{fig:reallocation_compare} illustrates the average performance degradation during periods of reallocation with network anomalies compared to normal periods. The latency and packet loss ratio both show significant increases, rising by factors of $4.49$ and $16$, respectively, while the bandwidth also experiences a decrease of $24\%$. The results above indicate that \textit{not all satellite reallocations affect the application level. However, when they do, the impact can significantly degrade the performance of Livecast applications}. Moreover, Fig. \ref{fig:dist_compare} illustrates the overall probability distribution during the normal period and the reallocation period with network anomalies. The figure highlights significant differences in both the mean and distribution of bandwidth and packet loss ratios between these two periods. This interleaved discrepancy, characterized by a minimum switch time of $15$ seconds, causes abrupt fluctuations in the LSNs, rendering general predictors ineffective, as shown in our subsequent evaluation.

Motivated by the above measurement and analysis, we advocate: \textbf{(i) Probabilistic prediction for LSNs.} Considering the challenges outlined above and the need to understand the pattern of UE-Sat reallocation to decide the suitable video quality and FEC ratio, we forecast the impact of future reallocation relying on historical network performance coupled with domain knowledge of reallocation characteristics. This method can bypass the analysis of complex and hidden satellite topology while remaining attuned to critical reallocation details. Additionally, we classify our data into distinct groups according to their association with network anomalies during reallocation and incorporate this information as a new input to the model. This modification enables the model to identify the presence of bimodal behavior in the time-series prediction task and generate more informative distribution predictions rather than just single-point predictions.

\textbf{(ii) Scheduler for distribution convolution.} Due to the high packet loss ratio during the reallocation period, adding parity packets further limits the already constrained uplink capacity, severely restricting the available goodput for video data in LSN's uplink. To address this, Variable Bitrate (VBR) encoding is necessary to achieve higher compression ratios compared to the general Constant Bitrate (CBR) encoding used in Livecast ingestion. As shown in our video set in \S \ref{sec:evalution}, VBR encoding provides a PSNR improvement of $2.35 \pm 3.27$ dB over CBR. In VBR, video quality is adjusted using the Constant Rate Factor (CRF), but the actual bitrate depends on texture complexity and motion speed. The variability in chunk sizes caused by VBR poses an additional challenge in determining the optimal FEC ratio, as it is closely tied to each video's chunk size. To address this, we introduce a CRF-bitrate distribution construction algorithm that analyzes real-time bitrate probability distributions and maps them to CRF values. Additionally, the video quality scheduler optimizes video parameters by jointly considering probabilistic predictions and bitrate distributions from video encoding.
\section{System Overview}
\label{sec:method}
\begin{figure}[t]
     \centering
     \includegraphics[width=\linewidth]{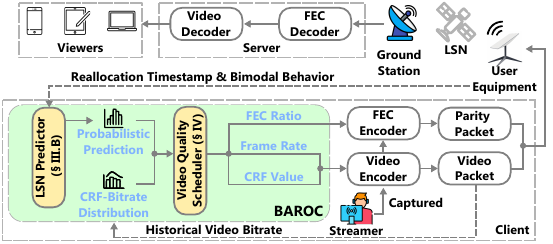}
     \caption{The overview of BAROC framework.}
     \label{fig:halo_overview}
     \vspace{-7mm}
\end{figure}

\subsection{BAROC overview}

In this section, we introduce BAROC, which addresses the challenges outlined in \S \ref{sec:intro} and incorporates the insights from the measurements detailed in \S \ref{sec:background}. BAROC consists of two main components: (i) a novel LSN predictor (\S \ref{sec:mtp_informer}) that provides probabilistic predictions of LSNs network status, and (ii) a video quality scheduler (\S \ref{sec:formulation}) designed to resolve the distribution convolution problem and determine the optimal video encoding settings. Figure \ref{fig:halo_overview} illustrates how BAROC fits into a Livecast ingestion scenario. The predictor first forecasts the network status distribution based on historical data and bimodal behavior distribution. Then, the scheduler considers both the CRF-bitrate distribution and the predicted network distribution to jointly determine the CRF value, frame rate, and FEC ratio. The client encodes the video and parity packets according to the selected video parameters and FEC ratio, and uploads them to the server via LSNs. The server utilizes the parity packets to recover any lost data, decodes the video, and distributes it to viewers. We will now provide a detailed description of the two components.

\subsection{MTP-Informer}
\label{sec:mtp_informer}

\begin{figure}[t]
     \centering
     \includegraphics[width=\linewidth]{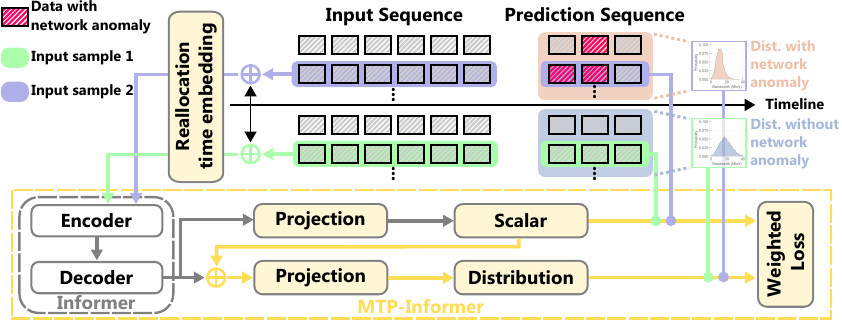}
     \caption{The overview of MTP-Informer.}
     \label{fig:mtp_overview}
     \vspace{-7mm}
\end{figure}

\textbf{Challenges of forecasting in LSNs}. Accurately predicting general network conditions is already challenging and resource-intensive \cite{tan2024accurate, lv2022lumos}. The additional complexities discussed in \S \ref{sec:intro}, such as reallocation events, bimodal behavior, and confidential network topology, further complicate forecasting in LSNs. Time-series forecasting models typically align predictions with more frequent historical points to minimize overall loss. However, in LSNs exhibiting bimodal behavior, infrequent but high-impact events demand greater attention, as highlighted in \S \ref{sec:background}. These bimodal behaviors often occur in an interleaved manner, with minimum intervals of $15$ seconds, and their actual impact on network performance remains unclear. Without precise information on these timestamps and intervals, general network predictors struggle to adapt to frequent behavior switching effectively.

\textbf{Network architecture.} The Transformer's capacity to capture long-range periodic patterns and dependencies \cite{vaswani2017attention, zhou2021informer, wu2021autoformer} makes it ideal for forecasting LSN status, including constant LEO satellite orbit periods and periodic reallocations. Additionally, its multi-head attention mechanism enhances feature extraction from diverse representation spaces. In detail, given a query matrix \( Q_i \in \mathbb{R}^{m \times d_q} \), a key matrix \( K_i \in \mathbb{R}^{m \times d_k} \), and a value matrix \( V_i \in \mathbb{R}^{m \times d_v} \) for the \(i\)-th head, where \(m\) is the sequence length, and \(d_q\), \(d_k\), and \(d_v\) are the dimensions of queries, keys, and values, respectively. The attention scores are computed using the scaled dot-product: $\text{Attention}(Q_i, K_i, V_i) = \text{softmax}\left(\frac{Q_i K_i^T}{\sqrt{d_k}}\right) V_i$. These attention scores are used to weight the values \(V_i\), and the outputs from all heads are concatenated and linearly transformed to obtain the final representation.

\textbf{Reallocation time embedding.} Based on the ability to learn different representations, we modified the model as follows: To make the new model aware of the reallocation timestamp and the intervals, we integrate one reallocation feature on the input data to act as an extra time embedding series as shown in Fig. \ref{fig:mtp_overview}. In detail, this series is a binary list where the reallocation second (e.g., 12th, 27th, 42th, and 57th) will be marked as $1$, and others as $0$. However, the end-to-end network performance doesn't always degrade when the reallocation feature is set to 1. To make the MTP-Informer aware of the probability of bimodal behaviors during the reallocation period, we introduce the multi-task structure to bridge this gap.

\textbf{Multi-task Structure.} As shown in Fig. \ref{fig:mtp_overview}, we extend the Informer model by dividing the prediction into two tasks: scalar prediction and distribution prediction. After the projector layer, the scalar matrix is of dimension \(\mathbb{R}^{bs \times \eta}\), and the distribution matrix is of dimension \(\mathbb{R}^{bs \times \eta \times \xi}\), where \(bs\) is the batch size, \(\eta\) is the prediction length, and \(\xi\) is the number of discrete distribution classes. The corresponding loss weights are then calculated from these outputs.

The scalar matrix's ground truth values can be directly obtained from the network trace, whereas the distribution matrix is unavailable. To address this, we classify each input sequence and its subsequent prediction sequence into two categories: those with network anomalies (packet loss ratio \(\geq 2\%\)) and those without, each linked to a specific distribution. The ground truth distribution matrix is then derived based on the category of the input sequence. For instance, if “green input sample 1” and its ground truth prediction sequence show no network anomalies, the distribution without network anomalies is used as the ground truth. 

In our initial exploration of MTP-Informer, we observed occasional conflicts between scalar and distribution predictions, such as instances where the distribution's mean deviated significantly from the scalar prediction. To address these discrepancies, and drawing inspiration from residual networks \cite{he2016deep}, we introduced a residual design by adding the scalar value before the projection layer in the distribution task. This scalar value serves as an approximate mean, guiding the distribution task toward faster convergence. This approach notably accelerates convergence and reduces weighted losses.

\textbf{Loss function.} We use $w^{sc}$ and $l^{sc}$ to represent the scalar value of bandwidth and packet loss ratio. We then denote the discrete distribution function of bandwidth and packet loss ratio classes as \(\mathcal{W}\) and \(\mathcal{L}\), respectively. Each item inside the distribution, $w \in Supp(\mathcal{W})$ and $l \in Supp(\mathcal{L})$, represents the probability of that class. The $Supp(\cdot)$ represents the possible values inside the distribution. Then the weighted loss for the model can be represented as:

\begin{equation}
\begin{gathered}
    \label{eq:model_loss}
    Loss = \epsilon_1 \sum_{t=1}^{\eta} F_{MSE}(w^{sc}_t, \hat{w}^{sc}_t) + \\
    \epsilon_2 \sum_{w \in Supp(\mathcal{W})} F_{CE}(w, \hat{w}) + \epsilon_2 \sum_{l \in Supp(\mathcal{L})} F_{CE}(l, \hat{l})
\end{gathered}
\end{equation}
where \(\eta\) denotes the length of the prediction sequence, \(F_{MSE}\) and \(F_{CE}\) represent the mean squared error (MSE) and cross-entropy loss respectively. The symbol with a hat indicates the corresponding ground truth value, and \(\epsilon_1\) and \(\epsilon_2\) are the weights assigned to the scalar loss and distribution loss. Overall, the MTP-Informer generates probabilistic predictions, providing more detailed information about potential low-frequency hazardous events in bimodal behavior to the subsequent video quality scheduler.
\section{Video Quality Scheduler}
\label{sec:formulation}

In this section, we formulate the distribution convolution problem arising from probabilistic prediction and the variance caused by VBR encoding. We propose a scheduler that takes into account both high- and low-frequency events in bimodal behavior, weighting their consequences to jointly determine the optimal video encoding parameters and FEC ratio.

\textbf{Problem formulation.} During the video delivery scenario, we model the time series as  $\mathcal{T} = \{1, 2 \cdots n, \cdots N\}$, the video bitrate and frame rate at each time $n$ are represented as $b_{n}$, and $\gamma_{n}$ respectively. Then we define the $\mathcal{F}_{n}$ as the frameset at time $n$ contains all the frames $f \in \mathcal{F}_{n}$, so that $||\mathcal{F}_{n}||=\gamma_{n}$. For each frame $f$, it will be packetized into multiple smaller packets, so that the packet can be transmitted through the Internet without further fragmentation. We assume the packet size is $12$ kb ($1500$ bytes) as it is the commonly used Maximum transmission unit (MTU) size\footnote{\url{https://en.wikipedia.org/wiki/Maximum_transmission_unit}}.

At each time $n$, the MTP-Informer will output the discrete probability distribution of bandwidth, and packet loss ratio, which will be denoted as $\{\mathcal{W}_{n}, \mathcal{L}_{n}\}$ respectively. In our formulation, we will use symbols $\mathcal{W}_{n}$, and $w_n$ to represent the distribution function and a sample value from the value space in time $n$. For instance, the Probability Mass Function of bandwidth $\mathcal{W}_{n}$ can be represented as:

\begin{equation}
\begin{gathered}
    \label{eq:time-frame-convert}
    P_{\mathcal{W}_{n}}(w_n), \text{with } \sum_{w_n \in Supp(\mathcal{W}_n)} P_{\mathcal{W}_{n}}(w_n) = 1, \forall n \in \mathcal{T}\\
\end{gathered}
\end{equation}

For the range and interval of each distribution\footnote{Refer to \S \ref{sec:experiment_setup} for the practical range and interval setup.}, the variables $w_n$ and $l_n$ have finite support domains defined as $w_n \in \{ I_w \times i \mid i \in \mathbb{Z}^{0+}, 0 \leq w_n \leq w_{max} \text{ kbps} \}$, and $l_n \in \{ I_l \times i \mid i \in \mathbb{Z}^{0+}, 0 \leq l_n \leq 1 \}$, where the $I_w$ and $I_l$ is the interval of bandwidth and packet loss ratio classes. The scheduler will output frame rate $\gamma_{n} \in \{\mathbb{Z}^{0+} | 0 \leq \gamma_{n} \leq \gamma_{max}\}$, CRF value $c_{n} \in \mathbb{C} = \{\mathbb{Z}^{0+} | c_1, c_2, \cdots, c_{max}\}$ and FEC ratio $\alpha_{n} \in \{\mathbb{R} | 0 \leq \alpha_{n} \leq 1\}$.

To quantify the number of packets, we use $u_f \in \mathbb{Z}^{0+}$ to represent the number of packetized packets for each frame $f$. Given the FEC ratio, the number of parity packets at frame $f$ can be expressed as \( p_f = \lceil \alpha_{n} u_f \rceil \). The ceiling function ensures that the resulting number of packets is both sufficient and an integer. To recover all lost packets, the number of total arrived packets should be larger or equal to the number of original packets $u_f$ \cite{rfc6865, begen2010rtp} after the packet loss event:

\begin{equation}
    \label{eq:fec_ratio}
        \lfloor (u_f + p_f) (1 - l_n) \rfloor \geq u_f, \forall f \in \mathcal{F}_{n}, l_n \in Supp(\mathcal{L}_{n})
\end{equation}

The floor function is used here to remove the fractional numbers introduced by the loss ratio. In the meanwhile, after the inclusion of parity packets, the throughput considering both original packets and parity packets also needs to be smaller than the available bandwidth to avoid congestion:

\begin{equation}
    \label{eq:bandwidth_limit}
    \begin{gathered}
        \sum_{f \in \mathcal{F}_{n}} (u_f + p_f)MTU \leq w_n, \forall n \in \mathcal{T}, w_n \in Supp(\mathcal{W}_{n})
    \end{gathered}
\end{equation}

The sum \((u_f + p_f)\) represents the total number of packets to be transmitted, and multiplying by the packet size \(MTU\) gives the frame size for frame \(f\). Given the two constraints above, we define the QoE on the receiver side at time \(n\) as:

\begin{equation}
    \label{eq:single_step_qoe}
    QoE_{n} = \omega_1 \gamma_{n} - \omega_2 c_{n} - \omega_3 |c_{n} - c_{n-1}|, \forall n \in \mathcal{T}
\end{equation}

The QoE aims to maximize the frame rate $\gamma_{n}$ and video quality, where a lower CRF value $c_{n}$ represents higher video quality. The $|c_{n} - c_{t_{n-1}}|$ is the penalty for quality changes. The $\{\omega_1, \omega_2, \omega_3\} \in \{\mathbb{R}| 0 \leq w \leq 1\}$ is the weights for different metrics. Based on the above constraints, the overall optimizing target can be defined as follows:  

\begin{equation}
\begin{gathered}
    \label{eq:qoe_target}
    \max_{\alpha_{n}, c_{n}}\sum_{n \in \mathcal{T}}QoE_{n}, \\
    s.t. \quad (\ref{eq:fec_ratio})-(\ref{eq:single_step_qoe})
\end{gathered}
\end{equation}

\textbf{Solution of distribution convolution problem.} The scheduler seeks to optimize the CRF value, FEC ratio, and frame rate to maximize QoE, considering the bandwidth and packet loss distributions. The bimodal behavior and resulting probability distributions complicate resolution using standard optimization methods. Furthermore, the interdependence of constraints (\ref{eq:fec_ratio}) and (\ref{eq:bandwidth_limit}) adds to the challenge. To address this, we employ a heuristic approach, first determining the optimal FEC ratio \(\alpha_{n}\) and then using our proposed algorithm to solve the distribution convolution problem to identify the optimal bitrate and frame rate. The FEC ratio \(\alpha_{n}\) is derived by relaxing constraint (\ref{eq:fec_ratio}) as follows:

\begin{equation}
    \label{eq:constrain_relaxing_2}
        (u_f + \lceil \alpha_{n} u_f \rceil) (1 - l_n) \geq u_f, \forall f \in \mathcal{F}_{n}, l_n \in Supp(\mathcal{L}_{n}),
\end{equation}

\begin{equation}
    \label{eq:constrain_relaxing_3}
        \lceil \alpha_{n} u_f \rceil \geq \frac{u_f}{1 - l_n} - u_f, \forall f \in \mathcal{F}_{n}, l_n \in Supp(\mathcal{L}_{n}),
\end{equation}

\begin{equation}
    \label{eq:constrain_relaxing_5}
        \alpha_n \geq \frac{1}{1 - l_n} - 1, l_n \in Supp(\mathcal{L}_{n}),
\end{equation}

\begin{equation}
    \label{eq:fec_distribution}
    \begin{gathered}
    P_{\mathcal{A}_{n}}(\alpha_n = \frac{1}{1-l_n} - 1) = P_{\mathcal{L}_{n}}(l_n), l_n \in Supp(\mathcal{L}_n)
    \end{gathered}
\end{equation}

Eq (\ref{eq:constrain_relaxing_2}) is derived from (\ref{eq:fec_ratio}) by replacing $p_f$ with $\lceil \alpha_{n} u_f \rceil$ and eliminating the floor function, where the relaxing is only feasible as the $u_f \in \mathbb{Z}^{0+}$. From (\ref{eq:constrain_relaxing_3}) to (\ref{eq:constrain_relaxing_5}), the ceiling function is relaxed and obtains a tighter lower bound for the FEC ratio shown as (\ref{eq:constrain_relaxing_5}). Then we can define the probability distribution of the FEC ratio $\mathcal{A}_{n}$ as (\ref{eq:fec_distribution}), which represents the minimal FEC ratio to cover each possible packet loss ratio in $Supp(\mathcal{L}_{n})$. With the distribution of the FEC ratio, we can then determine the available video bitrate using constraint (\ref{eq:bandwidth_limit}):

\begin{equation}
    \label{eq:bl_relaxing_1}
    \begin{gathered}
        \sum_{f \in \mathcal{F}_{n}} u_fMTU + \sum_{f \in \mathcal{F}_{n}}  \lceil \alpha_{n} u_f \rceil MTU \leq w_n,
    \end{gathered}
\end{equation}

\begin{equation}
    \label{eq:bl_relaxing_2}
    \begin{gathered}
        b_{n} + \sum_{f \in \mathcal{F}_{n}}  (\alpha_{n} u_f + 1) MTU \leq w_n,
    \end{gathered}
\end{equation}

\begin{equation}
    \label{eq:bl_relaxing_3}
    \begin{gathered}
        b_{n} + \gamma_{n} MTU + \alpha_{n} \sum_{f \in \mathcal{F}_{n}} u_f MTU \leq w_n,
    \end{gathered}
\end{equation}

\begin{equation}
    \label{eq:bl_relaxing_4}
    \begin{gathered}
        b_n \leq \frac{w_n - \gamma_n MTU}{\alpha_n + 1},
    \end{gathered}
\end{equation}
where $\quad \forall n \in \mathcal{T}, w_n \in Supp(\mathcal{W}_n),\alpha_n \in Supp(\mathcal{A}_n)$.

Constraint (\ref{eq:bl_relaxing_1}) is derived from (\ref{eq:bandwidth_limit}) by splitting the sum of the addition and replace $p_f$ with $\lceil \alpha_{n} u_f \rceil$. The sum of $u_fMTU$ for all frames within $\mathcal{F}_{n}$ is the bitrate $b_{n}$ at time $n$, we simplify the constraint to (\ref{eq:bl_relaxing_2}) with the ceiling function eliminated. Then, we decompose the sum using \(\sum_{f \in \mathcal{F}_{n}} MTU = ||\mathcal{F}_{n}||MTU\) to derive (\ref{eq:bl_relaxing_3}). Finally, the upper bound for the available bitrate \( b_{n} \) can be expressed as Eq (\ref{eq:bl_relaxing_4}). In real video encoding scenarios, achieving the constraint (\ref{eq:bl_relaxing_4}) may not always be feasible, especially when the available uplink is limited and the inclusion of parity packets. In such cases, it becomes necessary to proactively discard frames to maintain a balance between an acceptable frame rate and video quality. When the current frame rate $\gamma_n$ cannot satisfy the constraint in (\ref{eq:gamma_constraint}), a random P-frame within the current frameset $\mathcal{F}_{n}$ will be omitted until the constraint holds. Thus, the probability distribution of frame rate $\Gamma_n$ can be defined as:

\begin{equation}
    \label{eq:gamma_constraint}
    \begin{gathered}
        P_{\Gamma_{n}}(\gamma_n = \{max(\gamma_n) | w_n - \gamma_n MTU > 0\}) = P_{\mathcal{W}_{n}}(w_n), \\
        w_n \in Supp(\mathcal{W}_n), \gamma_n \in \{\mathbb{Z}^{0+} | 0 \leq \gamma_{n} \leq \gamma_{max}\}
    \end{gathered}
\end{equation}

To account for the effects of both high- and low-frequency events and their consequences based on the probability distribution, we calculate the Cartesian product of the values in the pair-wise set \(\{Supp(\mathcal{W}_{n}), Supp(\Gamma_{n})\}\) and \(Supp(\mathcal{A}_{n})\), along with their corresponding joint probabilities. This approach ensures that \(\mathcal{B}_{n}\) considers all combinations of possible bandwidth, FEC ratio, and frame rate, encompassing the outcomes of all possible events in the bimodal behavior:

\begin{equation}
    \label{eq:ava_bitrate_dist}
    \begin{gathered}
    P_{\mathcal{B}_{n}}(b_n = \frac{w_n - \gamma_n MTU}{\alpha_n + 1}) = P_{\mathcal{W}_{n}}(w_n) P_{\mathcal{A}_{n}}(\alpha_n), \\
    \{w_n, \gamma_n\} \in \{Supp(\mathcal{W}_n), Supp(\Gamma_n)\}, \alpha_n \in Supp(\mathcal{A}_n)
    \end{gathered}
\end{equation}

\begin{figure}[t]
\vspace{-5mm}
\begin{algorithm}[H]
\caption{CRF-Bitrate Distribution Construction Algorithm} 
\label{al:crf_bitrate}
\textbf{Input:} Selected CRF value $c_n$, Decoded bitrate $b_{n}$, Queue $Q$ with limited size, Current CRF-Bitrate distribution $\mathcal{M}$ \\
\textbf{Output:} New CRF-Bitrate distribution $\mathcal{M}$
\begin{algorithmic}[1]

\State{$Q[c_n].append(b_{n}$)}

\If {$||Q[c_n]|| < \theta_{minmal\_startup}$}
\State{$\mathcal{M}_{c_n} \leftarrow getDefaultDistribution(c_n)$}
\Else
\State{$\mathcal{M}_{c_n} \leftarrow fitMixtureDistribution(Q[c_n])$}
\EndIf
\State{\textbf{return} $\mathcal{M}$}

\end{algorithmic}
\end{algorithm} 
\vspace{-7mm}
\end{figure}

To figure out the relationship between CRF values and their corresponding encoded bitrates, we utilize a mapping table to record the historical video bitrates generated by each CRF value (See Algorithm \ref{al:crf_bitrate}). Given that the bitrate distribution can change with scene or subscene transitions in the video, we retain only the last $55$ seconds of bitrate data queue to adapt to the current video scenes \cite{cutting2016narrative}. In typical scenarios, the bitrate distribution is modeled using a mixture of Gaussian distributions. However, when the bitrate data queue is too short to accurately reflect the distribution, a pre-recorded default bitrate distribution is used initially. With both the distribution of available bitrate and the CRF-bitrate, the distribution convolution problem can be defined as follows:

\begin{equation}
\begin{gathered}
\label{eq:optimized_crf}
    c_{n} = \argmax_{c \in \mathbb{C}} \sum_{b_n \in Supp(\mathcal{B}_n)} P_{\mathcal{B}_{n}}(b_n) \cdot f_{\mathcal{M}_{c}}(m < b_n) \cdot \mathbb{E}(\mathcal{M}_c)
\end{gathered}
\end{equation}

The term \( f_{\mathcal{M}_{c}}(m < b_n) \) represents the probability that the current bandwidth \( b_n \) is sufficient for the CRF value \( c \). The expected value $\mathbb{E}(\mathcal{M}_c)$ serves as the reward for selecting the CRF value \( c \). Therefore, this equation aims to find the CRF value that maximizes the reward given all available bitrate combinations. Even though there are $||w_n|| \times ||l_n|| $ available bitrates in Eq (\ref{eq:optimized_crf}), the computations are independent and can be highly parallelized to accelerate the solving process. Given the optimal CRF value \( c_{n} \), we can then derive the most confident available bitrate \( b_n \) from Eq (\ref{eq:optimized_framerate_fec}). 

\begin{equation}
    \begin{gathered}
    \label{eq:optimized_framerate_fec}
            b_n = \argmax_{b_n \in Supp(\mathcal{B}_n)} P_{\mathcal{B}_{n}}(b_n) \cdot f_{\mathcal{M}_{c_{n}}}(m < b_n) \cdot \mathbb{E}(\mathcal{M}_{c_n}), 
    \end{gathered}
\end{equation}

This step does not require additional computation, as we can backtrack the computation process of Eq (\ref{eq:optimized_crf}) to obtain the arguments of the maxima. Consequently, we can also determine the optimal frame rate \( \gamma_n \) and FEC ratio \( \alpha_n \) that correspond to \( b_n \) in (\ref{eq:optimized_framerate_fec}). The above process describes how to achieve single-step optimization in Eq (\ref{eq:single_step_qoe}), while global optimization can be obtained using the dynamic programming approach detailed in Algorithm (\ref{al:qoe_optimizer}).

\begin{figure}[t]
\vspace{-5mm}
\begin{algorithm}[H]
\caption{Distribution Convolution Solving Algorithm}  
\label{al:qoe_optimizer}
\textbf{Input:} Next timestamp $n$, MTP-Informer $\phi(\cdot)$, Model Prediction length $\eta$\\
\textbf{Output:} Optimal decision $\gamma_{n}, c_{n}, \alpha_{n}$

\begin{algorithmic}[1]

\State{$\{\mathcal{W}_{n}, \mathcal{L}_{n}\}, \cdots \{\mathcal{W}_{n+\eta}, \mathcal{L}_{n+\eta}\} \leftarrow \phi(n)$}

\For{$\bar{n} \in [n, \cdots, n+\eta]$}
    \State{$c_{n} \leftarrow Eq_{(\ref{eq:constrain_relaxing_2})-(\ref{eq:optimized_crf})}(\mathcal{W}_{\bar{n}}, \mathcal{L}_{\bar{n}})$}
    \State{$\kappa \leftarrow \{\}$ \quad /* Empty DP-Table */}
    \For{$c \in [c_{n}, \cdots, c_{max}]$ } 
        \State{$\gamma, \alpha \leftarrow Eq_{(\ref{eq:optimized_framerate_fec})}(c)$ }
            \If{$\bar{n}=n$}  /* Initialize DP-Table */
                \State{$c_{n-1} \leftarrow $ Obtain from previous decision}
                \State{QoE $\leftarrow Eq_{(\ref{eq:single_step_qoe})}(\gamma, c, c_{n-1})$}
                \State{$\kappa_{\bar{n}}$.append(Node(QoE, $\gamma, c, \alpha$))}
            \Else
                \For{Node $\in \kappa_{\bar{n}-1}$}
                    \State{$c_{n-1} \leftarrow$ Node$.c$}
                    \State{Cumulative QoE $\leftarrow$ Node.QoE + $Eq_{(\ref{eq:single_step_qoe})}(\gamma, c, c_{n-1})$ }
                    \State{$\kappa_{\bar{n}}$.append(Node(Cumulative QoE, $\gamma, c, \alpha$))}
                \EndFor
            \EndIf
    \EndFor
\EndFor

\State{Node$_{max} \leftarrow $Find the Node with maximize QoE $\in \kappa_{n+\eta}$}
\State{$\gamma_{n}, c_{n}, \alpha_{n} \leftarrow$ Backtrack the start from the Node$_{max}$}
\State{\textbf{return} $\gamma_{n}, c_{n}, \alpha_{n}$}

\end{algorithmic}
\end{algorithm} 
\vspace{-7mm}
\end{figure}

\section{Evaluation}
\label{sec:evalution}

We evaluate BAROC using a trace-driven experiment utilizing real-world LSN traces with various videos. We compare BAROC with three representative methods in terms of multiple performance metrics from both video quality and FEC recovery perspectives. Additionally, we perform an ablation study to validate the effectiveness of the MTP-Informer compared to other time-series predictors, as well as to assess the contribution of each component within BAROC.

\subsection{Experiment Setup}
\label{sec:experiment_setup}

\textbf{Baselines and evaluation metrics.} We compare BAROC with three representative Livecast video recovery frameworks LightFEC \cite{hu2021lightfec}, R-FEC \cite{lee2022r}, and FBRA \cite{nagy2014congestion}.

\begin{itemize}[left=0pt]
    \item LightFEC: A network-adaptive FEC system employed LSTM with a clustering algorithm to learn the characteristics of different networks.
    \item R-FEC: A reinforcement learning based framework for video and FEC bitrate decision. The RL model will output the optimal FEC ratio and bitrate according to the historical network status.
    \item FBRA: A FEC-based Rate Adaptation algorithm designed for \emph{WebRTC}. The algorithm utilizes a state machine to determine the optimal bitrate and FEC ratio.
\end{itemize} 
We consider four evaluation metrics as follows: i) PSNR between received video and source video, ii) Frame rate at the receiver side after FEC recovery, iii) recovery ratio, the number of recovered packets divided by the number of lost packets, and iv) parity packet utility, the number of recovered packets divided by the number of parity packets.

\textbf{Packet-level video and LSN traces.} Our video set includes three 10-minute videos with a resolution of $1920 \times 1080$ at $60$ FPS: Chatting \cite{TwtichVideoChatting}, Gaming \cite{TwtichVideoGaming}, and Sports \cite{YoutubeVideoSport}. These scenarios represent typical Livecast use cases, covering both dynamic scenes, such as a running basketball player, and static ones, like facial expressions or slight body movements. For VBR encoding, we adjust the CRF value to modify video quality, achieving higher frame-level compression compared to directly altering the quantization parameter\footnote{\url{https://slhck.info/video/2017/02/24/crf-guide.html}}. Videos are encoded with CRF values \(\{26, 31, 36, 41, 46, 51\}\), representing six quality levels, producing bitrates from $500$ kbps to $15,000$ kbps, matching LSN's uplink capacity. The \( w_{max} \) for bandwidth classes in MTP-Informer is also set to $15,000$ kbps. Intervals \( I_w \) and \( I_i \) are $500$ kbps and $0.02$, respectively, balancing computational cost and data granularity. For baselines that only adjust video bitrate, we encode the source video using CBR, adjusting the bitrate to match the file size generated by VBR. The difference in file size between CBR and VBR videos is less than $0.1\%$.

To capture detailed network behavior during Livecast at packet-level granularity, we use \emph{FFmpeg} to stream the videos to a localhost media server and record the packet-level traffic. Modifications to the \emph{rtpenc.c} file enable the logging of RTP packet details, including sequence number, frame presentation time, packet size, and associated frame number. We employ \emph{RTSP} as the application-level protocol for video delivery instead of mainstream options like \emph{WebRTC}, as it is easier to modify and lacks a built-in packet recovery mechanism. This ensures a pure video data flow without redundant packets. Moreover, since both \emph{RTSP} and \emph{WebRTC} use \emph{RTP} for multimedia delivery, \emph{RTSP} serves as a suitable representation of real-world packet flow in Livecast scenarios.

The LSN dataset used for model training is collected from a \emph{Starlink} UE located around the West Coast of the American continent. This UE has an unobstructed view of the sky, and its performance is similar to that observed in multiple previous measurement studies \cite{ma2023network, michel2022first, kassem2022browser, mohan2024multifaceted, zhao2023realtime}. We collected a total of $75$ hours of LSN network data using \emph{iperf3} and \emph{ping} to an AWS server. This data includes three network metrics: latency, server received bandwidth, and packet loss ratio, all recorded at a granularity of 1 second. The dataset was then divided into training, validation, and test sets, with each set containing $70\%$, $20\%$, and $10\%$ of the data, respectively.

\textbf{Experiment implementation.} The MTP-Informer is implemented using \emph{PyTorch} based on the original Informer implementation\footnote{\url{https://github.com/zhouhaoyi/Informer2020}}. To accurately emulate an environment that includes packet-level delivery, FEC recovery, and video playback, we developed a simulator in \emph{Python} consisting of approximately 1,700 lines of code. This simulator models packet-level behaviors such as network congestion, packet loss, and packet timeouts, applies FEC recovery on the receiving end, and evaluates their impact on the received video statistics. It is worth mentioning that BAROC can be deployed either on the streamer side or the server side, depending on the available computation capacity. The server, capable of monitoring the network status on the streamer's upload link, can predict conditions, determine the optimal CRF value and FEC ratio, and package these decisions into an RTCP receiver report packet to send back to the streamer.

\subsection{BAROC Comparison}

\begin{figure*}[t]
     \centering
     \includegraphics[width=\linewidth]{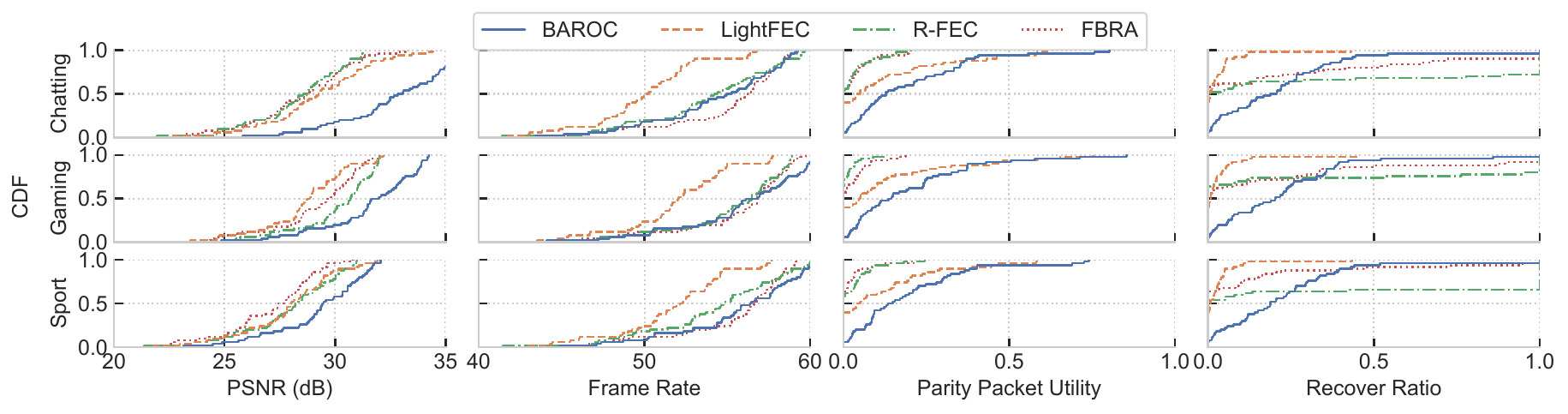}
     \caption{The CDFs of different baselines on different videos.}
     \label{fig:CDF_compare}
     \vspace{-5mm}
\end{figure*}

The experiment will utilize only the network trace from the aforementioned test set, which is approximately $7.5$ hours long. For each baseline method, the video will be repeated $50$ times using the test data until it is exhausted. 

\textbf{Video quality comparison.} From Fig. \ref{fig:CDF_compare}, BAROC outperforms other baselines in terms of PSNR, with an average improvement of $1.95$ dB and a maximum improvement of $3.44$ dB in the Chatting video. BAROC performs particularly well with relatively static scenes, as VBR encoding can achieve a better compression ratio using the appropriately selected CRF value. In videos with frequent scene changes, such as Sport, BAROC still maintains a minimum advantage of $1.27$ dB compared to other baselines, contributed by the awareness of bimodal behavior and sudden bandwidth degradation. On the other hand, the state machine used by FBRA struggles to adapt to the frequent behavior switches in LSNs, and its static switching rules cannot accommodate the bimodal behaviors in LSNs, resulting in the worst PSNR value. 

In terms of video frame rate, BAROC achieves a maximum improvement of 4.02 FPS compared to LightFEC and an average improvement of 1.53 FPS. FBRA shows a slight advantage because it tends to stay in the "Down" state during frequent packet loss and bandwidth decreases in LSNs, resulting in the long-term selection of the lowest bitrate, which yields a higher frame rate after FEC recovery. The LightFEC performs significantly worse in terms of frame rate compared to its performance on PSNR. The LSTM in LightFEC often overestimates the available bandwidth during the switching between bimodal behaviors, resulting in excessively high bitrates selection and bursts of frame losses. This phenomenon indicates that a general time-series predictor is unlikely to accurately account for bimodal behavior without specific network modifications and additional reallocation information.

\textbf{Recovery efficiency comparison.} R-FEC and MTP-Informer employ different strategies for selecting the FEC ratio. R-FEC prioritizes mitigating packet loss by aggressively filling the available bandwidth with parity packets. Without information on reallocation timestamps and bimodal behavior distribution, R-FEC maximizes frame rate by consistently sending saturated parity packets, as its reward function considers both video bitrate and frame rate \cite{lee2022r}. While this approach minimizes sudden bursty packet loss during reallocations, it compromises video quality. In contrast, BAROC increases the FEC ratio only when it predicts a bimodal behavior switch during reallocations. As a result, R-FEC achieves the highest recovery ratio but with very low parity packet utility. BAROC, however, improves parity packet utility by $6.08\%$ to $353\%$ compared to other baselines while maintaining a superior recovery ratio over LighFEC and FBRA. Notably, the shape of the CDF curves remains consistent across videos, as the same network trace clips are used. Recovery efficiency metrics are more sensitive to network conditions than to video content.

\subsection{Ablation Study}

\begin{figure}[t]
     \centering
     \includegraphics[width=\linewidth]{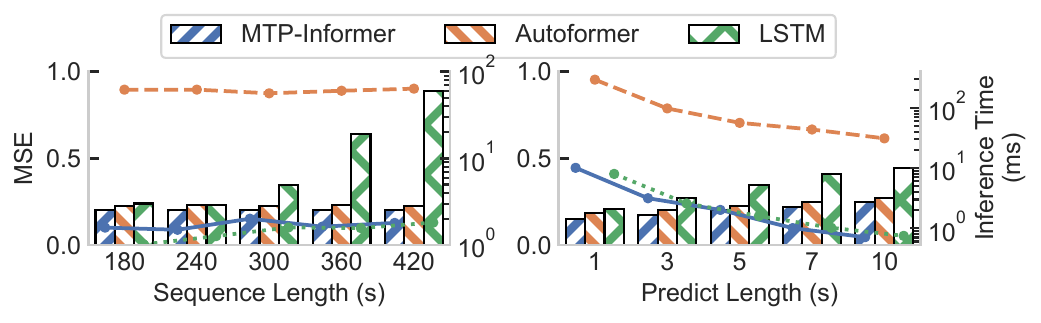}
     \caption{The trade-off between model accuracy and model inference time. The bar plot represents the MSE and the line plot reflects the inference time.}
     \label{fig:model_length_compare}
     \vspace{-5mm}
\end{figure}

We first isolate the prediction models to focus solely on investigating the prediction accuracy of the MTP-Informer, compared to two representative time-series predictors, Autoformer \cite{wu2021autoformer} and LSTM. Then, we separate the components in BAROC to thoroughly understand the improvements contributed by each component.

\textbf{Isolation of prediction models.} To evaluate the model prediction accuracy under the LSNs, we first evaluate the model accuracy with different model hyperparameters. We evaluate the MSE between different input lengths and prediction lengths for each model. Based on our initial exploration of optimal lengths, we empirically shrink the optimal range of input length to $\{180, 240, 300, 360, 420\}$, and the range of prediction length to $\{1, 3, 5, 7, 10\}$ seconds. To simplify the parameter combinations, we adhere to the single variable principle to test the effect of each parameter individually. 

All models are trained using a \emph{RTX A5000} GPU, and the results are shown in Fig \ref{fig:model_length_compare}. Comparing different input lengths, an input length of $180$ seconds already provides sufficient observation space for the two Transformer models, where the MSE is nearly the same as the input length growth. Yet, extending it further could complicate the learning process for another model, as indicated by degradation from LSTM. As the prediction length increases, all models show a slight rise in MSE due to the challenges of forecasting the distant future. Longer prediction lengths allow more time for video quality scheduling, which is beneficial for devices with computational constraints, such as mobile devices. For inference time, except Autoformer, all models maintain an inference time of less than 10 ms, which is acceptable for making real-time decisions.

\begin{figure}[t]
     \centering
     \includegraphics[width=\linewidth]{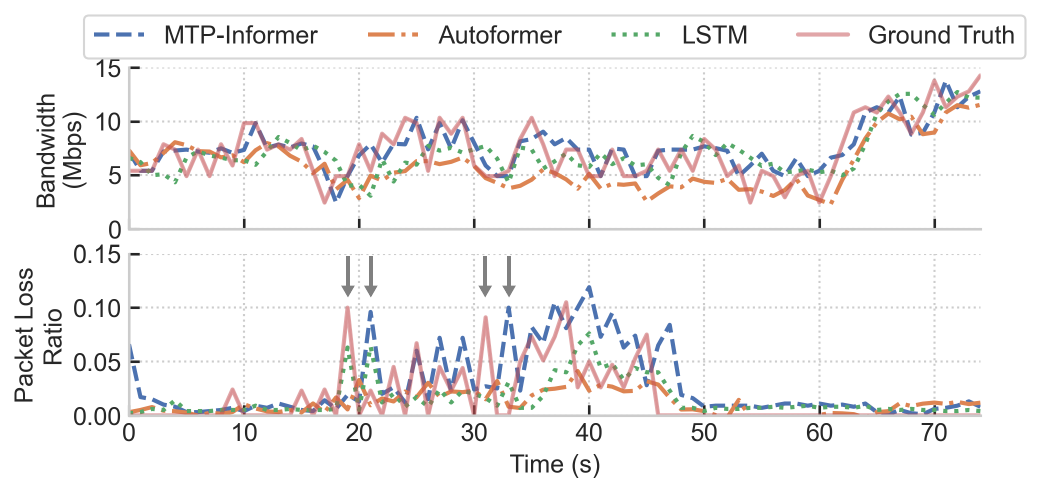}
     \caption{The prediction result of all models with its optimal input and prediction length compared to the ground truth.}
     \label{fig:pred_ground_compare}
     \vspace{-5mm}
\end{figure}

The MSE can only represent the overall quality of prediction, we also conducted a fine-grained comparison to reveal the model performance under LSNs's bimodal behaviors. As shown in Fig. \ref{fig:pred_ground_compare}, the MTP-Informer more closely matches the ground truth data compared to the other two models, in terms of both bandwidth and packet loss ratio. However, the single-point prediction from models still experiences errors when the packet loss ratio suddenly decreases or increases, as indicated by the arrows in Fig. \ref{fig:pred_ground_compare}. The probabilistic prediction, integrated with the video quality scheduler, helps mitigate the impact of unexpected packet loss events, with its effectiveness demonstrated in later isolation experiments. It is worth mentioning that to ensure a fair comparison, only the scalar value in the MTP-Informer is used to calculate loss in the experiments above, as the other models do not output distributions.

\begin{figure}[t]
     \centering
     \includegraphics[width=\linewidth]{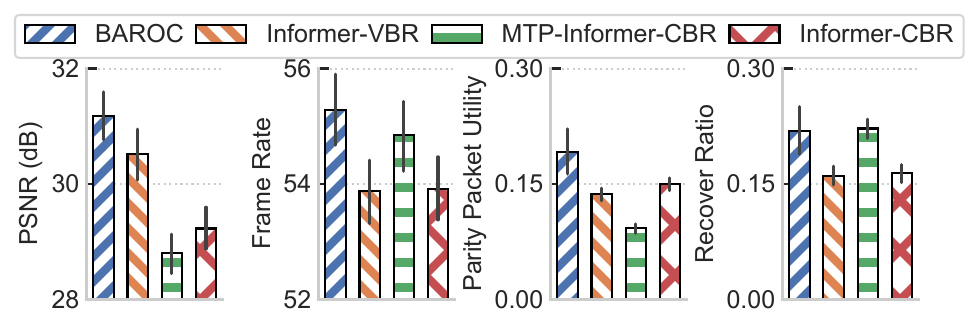}
     \caption{The comparison between BAROC and ablation variants.}
     \label{fig:ablation}
     \vspace{-5mm}
\end{figure}

\textbf{Isolation of BAROC components.} We separate the probabilistic prediction and video quality scheduler components in BAROC, resulting in the following ablation variants: i) Informer-VBR: The original Informer predictor combined with video quality scheduler with VBR encoding. To integrate it within our scheduler, we replace \(b_n\) in Eq (\ref{eq:optimized_crf}) with the single-point prediction, and a probability of  $100\%$, to obtain the optimal CRF value. A similar method is used to determine the FEC ratio from Eq (\ref{eq:constrain_relaxing_5}). ii) MTP-Informer-CBR: The MTP-Informer using the CBR encoding method. We use the expected value of the output probability distribution as the single-point prediction for bandwidth and loss ratio, which is then used to select the optimal video quality and FEC ratio. iii) Informer-CBR: The Informer model with CBR encoding is used as the comparison baseline.

As shown in Fig. \ref{fig:ablation}, BAROC surpasses all ablation variants in PSNR, with improvements ranging from $0.65$ dB to $2.36$ dB. The MTP-Informer-CBR variant performs the worst, as relying on the expected value of the distribution often results in inaccurate bandwidth and packet loss predictions, underscoring the importance of addressing the distribution convolution problem. Informer-VBR demonstrates notable gains over Informer-CBR, leveraging VBR encoding and the proposed video quality scheduler. However, neither Informer variant accounts for the bimodal behavior during satellite reallocations, leading to frame losses during sudden packet loss or bandwidth degradation. In terms of recovery efficiency, while MTP-Informer-CBR achieves comparable recovery ratios to BAROC due to its awareness of bimodal behavior, BAROC consistently outperforms all other ablation variants.

\textbf{Decision time comparison.} We evaluate the decision time of each model-based framework to demonstrate the feasibility of BAROC in Livecast ingestion. The FBRA is excluded in this comparison as the rule-based method is lightweight and has a decision time of less than $1$ ms in our experiment. Table. \ref{table:decision_time} shows the single decision time for all models including the original Informer as a baseline. The decision time includes both model inference time and the video quality scheduling time, so overhead of the proposed algorithm is around $4.18$ ms compared to the Informer model. Additionally, the model inference time could be even shorter with a more powerful GPU, as the \emph{A5000} is not among the latest GPUs.

\begin{table}[t]
\centering
\begin{tabular}{ccccc}
\hline
 & BAROC & LightFEC & R-FEC & Informer \\ \hline
Decision Time (ms) & 16.10 & 10.49 & 14.22 & 11.92 \\ 
Standard Deviation & 0.36 & 1.07 & 1.47 & 0.28 \\ \hline
\end{tabular}
\caption{Decision time for all model-based frameworks with Informer as a baseline of the model inference time.}
\vspace{-5mm}
\label{table:decision_time}
\end{table}
\section{Related Work}

\textbf{Measurements and optimization in LSNs.} In recent years, there has been significant exploration of LSNs across various domains. This exploration spans theoretical analyses \cite{hauri2020internet, handley2018delay, chen2021time}, early-stage measurements \cite{khalife2021first, tregloan2021optical, tregloan2020first}, and system design considerations \cite{bhattacherjee2019network, lai2023achieving, singh2021community, chen2022delay, li2022optimized}. Recently, general users and measurement studies \cite{ma2023network, michel2022first, kassem2022browser, zhao2023realtime} have recognized the real-world challenges in current LSN implementations. Great efforts have been made to address these challenges from both transport and application level \cite{cao2023satcp, wu2024accelerating, li2024satguard, zhao2024low, fang2024robust, zhang2024starstream}. However, issues such as limited uplink capacity and fluctuating performance during satellite reallocation remain unresolved for Livecast ingestion scenarios. This work addresses these challenges and complements the existing efforts by focusing on the Livecast ingestion application.

\textbf{Livecast video recovery.} With the increasing demand for Livecast applications, determining the appropriate FEC ratio has become more challenging by solely using the early stages FEC schemes like parity codes \cite{begen2010rtp} and Reed-Solomon codes \cite{rfc6865}. Recent approaches like Ivory \cite{emara2022ivory}, R-FEC \cite{lee2022r}, and LightFEC \cite{hu2021lightfec} utilized techniques such as LSTM and Reinforcement Learning to predict future network packet loss and optimize the FEC ratio. Other works like \cite{ray2022prism, nagy2014congestion, rudow2023online, rudow2023tambur, fong2019low, chenggrace, shen2023sja} focused on recovering bursts of packet loss or utilizing Neural Codecs for frame-level recovery. In general, those solutions are either expensive for Livecast ingestion or they need sufficient bandwidth for redundancy transmission, and their prediction mechanisms are not aware of the unique bimodal behavior caused by satellite reallocation. Therefore, our work complements this gap in the Livecast video recovery framework under LSNs.

\section{Conclusion}
In conclusion, this paper presented BAROC, a framework for bimodal behavior-aware loss recovery in LSNs, tailored for Livecast ingestion applications. Key contributions include: i) identifying bimodal behavior during satellite reallocation in LSNs, ii) introducing the MTP-Informer to account for bimodal behavior and fluctuated uplink bandwidth, and iii) formulating the distribution convolution problem to integrate prediction probabilities with bitrate distributions and solved through a video quality scheduler. Experiments with real-world LSN traces and diverse videos have demonstrated BAROC's effectiveness compared to state-of-the-art recovery methods. This work only focused on a specific portion of large images of multimedia applications under LSNs. In our future work, we aim to leverage the processing and caching capabilities of LEO satellites alongside the global coverage of LSN providers. By enabling in-orbit processing of multimedia content and minimizing unnecessary interconnection point routing, we can further enhance transmission efficiency and reduce latency.

\balance
\bibliographystyle{IEEEtran}
\bibliography{reference.bib}

\end{document}